\documentclass[]{aa} 

\usepackage{txfonts}
\usepackage{longtable}  
\usepackage{rotating}
\usepackage{natbib}
\usepackage{graphics}
\usepackage{graphicx}
\usepackage{bmpsize}
\usepackage{psfrag}
\usepackage{amssymb}
\usepackage{multicol}
\usepackage{pdflscape}
\usepackage{rotating}
\bibpunct{(}{)}{;}{a}{}{,}
\def\Teff{\ensuremath{T_{\mathrm{eff}}}}
\def\logTeff{\ensuremath{\log T_{\mathrm{eff}}}}
\def\logg{\ensuremath{\log g}}

\def\vsini{\ensuremath{{\upsilon}\sin i}}
\def\kms{$\mathrm{km\,s}^{-1}$}

\def\logl{\ensuremath{\log L/L_{\odot}}}

\begin{document}

\title{Orbital parameters and evolutionary status of the highly
peculiar binary system HD 66051}

\author{E.~Paunzen\inst{1} 
\and M.~Fedurco\inst{2}
\and K.G.~He{\l}miniak\inst{3}
\and O.I.~Pintado\inst{4}
\and F.-J.~Hambsch\inst{5,6,7}
\and S.~H{\"u}mmerich\inst{5,6}
\and E.~Niemczura\inst{8}
\and K.~Bernhard\inst{5,6}
\and M.~Konacki\inst{3}
\and S.~Hubrig\inst{9}
\and L.~Fraga\inst{10}}
\institute{Department of Theoretical Physics and Astrophysics, Masaryk University,
Kotl\'a\v{r}sk\'a 2, CZ-611\,37 Brno, Czech Republic
\email{epaunzen@physics.muni.cz}
\and{Faculty of Science, P. J. {\v S}af{\'a}rik University, Park Angelinum 9, Ko{\v s}ice 040\,01, Slovak Republic}
\and{Nicolaus Copernicus Astronomical Center, Polish Academy of Sciences, ul. Rabia\'{n}ska 8, PL-87-100 Toru\'{n}, Poland}
\and{Centro de Tecnolog{\'i}a Disruptiva, Universidad de San Pablo Tucum{\'a}n, San Pablo, 
Tucum{\'a}n, Argentina}
\and{American Association of Variable Star Observers (AAVSO), 49 Bay State Rd, Cambridge, MA, 02138, USA}
\and{Bundesdeutsche Arbeitsgemeinschaft für Ver{\"a}nderliche Sterne e.V. (BAV), D-12169, Berlin, Germany}
\and{Vereniging Voor Sterrenkunde (VVS), Brugge, BE-8000, Belgium}
\and{Instytut Astronomiczny, Uniwersytet Wroc{\l}awski, Kopernika 11, PL-51-622, Wroc{\l}aw, Poland}
\and{Leibniz-Institut für Astrophysik Potsdam (AIP), An der Sternwarte 16, D-14482 Potsdam, Germany}
\and{Laborat{\'o}rio Nacional de Astrof{\'i}sica LNA/MCTI, R. Estados Unidos, 154, Itajub{\'a} 37504-364, MG, Brazil}
}

   \date{} 
  \abstract
{The spectroscopic binary system HD\,66051 (V414 Pup) consists of a highly peculiar CP3 (HgMn) star and an A-type component. It also shows out-of-eclipse variability that is due to chemical spots. This combination allows the derivation of tight constraints for the testing of time-dependent diffusion models.}
{We aim at deriving astrophysical parameters, information on age, and an orbital solution of the system.}
{We analysed radial velocity and photometric data using two different methods to determine astrophysical parameters and the orbit of the system. Appropriate isochrones were used to derive the age of the system.}
{The orbital solution and the estimates from the isochrones are in excellent agreement with the estimates from a prior spectroscopic study. The system is very close to the zero-age main sequence and younger than 120\,Myr.}  
{HD\,66051 is a most important spectroscopic binary system that can be used to test the predictions of the diffusion theory explaining the peculiar surface abundances of CP3 stars.}

\keywords{binaries: eclipsing  -- stars: chemically peculiar -- stars individual: HD 66051 -- techniques: photometric}

\titlerunning{Highly peculiar binary system HD 66051}
\authorrunning{Paunzen et al.}
\maketitle

\section{Introduction} \label{introduction}

The classical chemically peculiar (CP) stars are early B- to early F-type objects that comprise about 10\% of upper main-sequence stars. They are characterized by peculiar atmospheric compositions that deviate significantly from the solar pattern. Most CP stars are characterized by large excesses (up to several orders of magnitude) of heavy elements such as Si, Eu, Hg, or the rare-earth elements \citep{Preston74}. The observed chemical peculiarities are thought to be caused by atomic diffusion \citep{Michaud76}. In this model, radiative levitation and gravitational settling operate in the stellar atmospheres of slowly rotating stars. Without significant convection, most elements sink under the influence of gravity; however, those with absorption lines near the local flux maximum are radiatively driven outward.

The CP stars are commonly divided into four groups \citep{Preston74}: CP1 stars (the metallic-line or Am/Fm stars), CP2 stars (the magnetic Bp/Ap stars), CP3 stars (the mercury-manganese or HgMn stars), and CP4 stars (the He-weak stars). Whereas the CP2/4 stars are characterized by globally organized magnetic fields, the CP1/3 stars are generally considered to be non-magnetic or only weakly magnetic objects \citep{Hubrig12}.

The CP3 stars, which are relevant to the present investigation, exhibit unusually strong enhancements of Hg and Mn (up to 6 and 3 dex, respectively), increased abundances of elements such as P, Y, Sr, and Xe, and depletions of other elements such as He, Ni, or Al \citep{Castelli04}. Although CP3 stars show individualistic abundance patterns, the strength of the observed overabundances generally increases with atomic number \citep{Ghazaryan16}.

The inhomogeneous surface distribution of elements (chemical spots) has been well documented spectroscopically for this type of stars \citep{Briquet10,Hubrig10,Maka11,Korh13}. Furthermore, evidence for secular evolution of these structures has been presented \citep{Koch07}. The question whether these chemical spots lead to photometric variability has not been tackled in detail so
far. Arguments have been presented in favour of both rotational modulation via the redistribution of flux in the spots and pulsational variability in CP3 stars (see e.g. the discussion in \citealt{Huemm18}). However, recent studies strongly suggest that rotational modulation is at the root of the observed photometric variations in at least some CP3 stars \citep{Morel14,Strass17,White17,Huemm18}.

A case similar to that of our target star was presented by \citet{Strass17}, who identified complex out-of-eclipse variability in the eclipsing double-lined spectroscopic binary system HSS\,348. It was found that at least the primary component is a CP3 star, and the authors concluded that the complex but stable out-of-eclipse variability is due to rotational modulation caused by an inhomogeneous surface distribution of elements.

It has yet to be fully understood why time-dependent diffusion processes are creating significantly different atmospheric compositions in CP3 stars of similar temperature \citep{Urpin15}. To investigate this topic in more detail and to answer the question of which other processes (if any) are at work in the creation of CP3 star anomalies in addition to atomic diffusion, corresponding data for stars with precise age determination are needed.

Here, we present an important step in this direction with the calculation of orbital parameters and age determination of the eclipsing binary system HD\,66051 (V414 Pup). \citet{Niem17} published detailed abundances of this binary system, which consists of a highly peculiar CP3 primary and an A-type secondary and boasts an orbital period of about 4.75\,d. Out-of-eclipse variability with the same period was identified and attributed to rotational modulation caused by chemical spots. The unique configuration of HD\,66051 allows studying phenomena such as atmospheric structure, mass transfer, magnetic fields, photometric variability, and the origin of chemical anomalies observed in CP3 stars and related objects \citep{Niem17}. Furthermore, the study of the system during eclipses allows precise Zeeman–Doppler imaging of the components \citep{Hubrig95}.

From photometric and radial velocity (RV) measurements, we have derived orbital parameters using two different approaches. We conclude that HD\,66051 is a young binary system close to the zero-age main sequence, and the primary CP3 component indeed shows a spotty surface. We consider this object a keystone for analysing time-dependent diffusion at very early stages of the
main-sequence evolution.

\section{Observations and reductions} \label{observations}

Photometric observations were acquired at the Remote Observatory Atacama Desert (ROAD) with an Orion Optics, UK Optimized Dall Kirkham 406/6.8 telescope and an FLI 16803 CCD camera. Data were obtained through Astrodon Photometric $BVI_{\mathrm{C}}$ filters. A total number of 2\,803, 2\,776, and 2\,271 ($B$, $V$, $I_{\mathrm{C}}$) observations were taken during a time span of 60, 60, and 37 days, respectively. The reductions were performed with the MAXIM DL program\footnote{http://www.diffractionlimited.com/} and the determination of magnitudes using the LesvePhotometry program\footnote{http://www.dppobservatory.net/}. These data have been presented in \citet{Niem17}, although without further analysis and model fitting.

For the determination of RVs, the following data have been used:
\begin{itemize}
\item Five spectra from the HIgh-Dispersion Echelle Spectrograph (HIDES) at the 1.88m telescope of the Okayama Astrophysical Observatory (OAO) in Japan, R\,$\sim$\,50\,000, 4090\AA\ to 7520\AA, taken during two runs between February and April 2017. The in-eclipse observation procured with the same instrument and employed by \citet{Niem17} was taken during the same time, but we did not use it here.
\item Six spectra from the REOSC spectrograph at the 2.15m telescope of CASLEO in Argentina, R\,$\sim$\,12\,000, 4150\AA\ to 5750\AA, taken during two nights in May 2017.
\item One spectrum from the High Accuracy Radial velocity Planet Searcher (HARPS) spectrograph at the ESO La Silla 3.6m telescope in Chile, R\,$\sim$\,110\,000, 3900\AA\ to 6900\AA, which was procured from the ESO archive (Prog. ID: 077.D-0085). This is the spectrum that \citet{Niem17} primarily based their analysis on.
\end{itemize}

The RVs were measured with our own implementation of the two-spectra cross-correlation function (CCF) technique TODCOR \citep{Zuck94}, with synthetic spectra calculated for the atmospheric parameters and abundances obtained by \citet{Niem17}: \Teff\,=\,12\,500~K, \logg\,=\,4.0, and \vsini\,=\,27~\kms\ for the primary, and \Teff\,=\,8\,000~K, \logg\,=\,4.0, and \vsini\,=\,18~\kms\ for the secondary star. The synthetic spectra were calculated with the ATLAS\,9 and SYNTHE codes \citep{Kuru05}, ported to GNU/Linux by \citet{Sbor05}. Atomic data were taken from \citet{Castelli04} and supplemented for the second and third spectra of the lanthanides with data taken from the Vienna Atomic Line Database \citep[VALD,][]{Kupka99} that have originally been presented in the Data on Rare Earths At Mons University database \citep[DREAM,][]{Biem99}. Single measurement errors were calculated using a bootstrap approach, as in \citet{Helm12}.

\begin{table}
\caption{Orbital elements and astrophysical parameters of HD\,66051. The abbreviations are explained in the text.}
\label{orbital_elements}
\begin{center}
\begin{tabular}{lc}
\hline
Parameter & Value\\
\hline
$P_{\mathrm{orb}}$\,(d) & 4.749218 \\
$T_{\rm 0}$ (HJD)& 2452167.867 \\ 
$K_{1}$\,(\kms)& 77.24(60) \\
$K_{2}$\,(\kms)& 137.97(90) \\
$e$& 0.0 \\
$\omega$\,(deg)& 0.0 \\
$\gamma$\,(\kms) & $-$1.6(3) \\
SMA\,(R$_{\odot}$) & 20.370(4) \\                              
$i$\,(deg) & 84.7(1)\\                              
$T_{\mathrm{eff,1}}$\,(K)& 12\,500  \\                              
$T_{\mathrm{eff,2}}$\,(K) & 8975(480) \\                          
$q$       & 0.56(3) \\                             
$\Omega_{\rm 1}$ & 8.50(9) \\                             
$\Omega_{\rm 2}$ & 8.22(7) \\ 
$M_{\mathrm{Bol,1}}$\,(mag) & $-$0.62(16) \\                    
$M_{\mathrm{Bol,2}}$\,(mag) & +1.81(12) \\ 
$M_{\rm 1}$\,(M$_{\odot}$) &    3.23(22) \\
$M_{\rm 2}$\,(M$_{\odot}$) &    1.81(13) \\    
$R_{\rm 1}$\,(R$_{\odot}$) & 2.58(13) \\                    
$R_{\rm 2}$\,(R$_{\odot}$) & 1.61(10) \\                    
\hline   
\end{tabular}    
\end{center}                                      
\end{table}

\section{Results} \label{analysis}

Based on the available measurements, we calculated the orbital parameters and 
investigated the location of the components within the Hertzsprung-Russell
diagram.

\subsection{Analysis of the orbital parameters} \label{analysis_phoebe}

We obtained orbital parameters using two widely different methods, whose results are in excellent agreement. The final results are presented in Table \ref{orbital_elements}.

The $rms$ of the fit is significantly larger than the individual RV errors, which suggests that they were under-estimated and did not include possible systematics, which often happens for fast-rotating or spotted stars. We therefore introduced (added in quadrature) a jitter term of 0.42 and 0.82~\kms\ for the primary and secondary, respectively. This term compensates for systematics coming from the rotational broadening or the surface inhomogeneities, for instance, and it reflects the spread of the RVs around the orbital fit. Moreover, since REOSC is the least precise spectrograph of the three instruments involved, its measurements are likely affected by instrumental systematic errors as well. To compensate for this, we have added an additional term of 2.5~\kms\ to all REOSC RV errors. With all modified uncertainties, we reached a reduced $\chi^2$ of the fit very close to 1, ensuring that the formal uncertainties of the fitted quantities are not underestimated. Finally, we checked the influence of other systematics (such as poor sampling, low number of measurements, and pulsations) on the uncertainties of the output parameters by running a bootstrap analysis with 10\,000 iterations. We found that they dominate the fit, and adopted the bootstrap parameter errors as the final ones. Notably, the relative mass uncertainty is still at a very good level of $\sim$1.5~\%. 

As a first approach, we employed the code V2FIT \citep{Kona10}, which uses the Levenberg-Marquard minimization scheme. The orbital period was held fixed to the value of 4.749218~d \citep{Niem17}. We fitted the two velocity semi-amplitudes $K_{\mathrm{1,2}}$, the systemic velocity (velocity of the centre of mass) $\gamma$, and the zero-phase $T_{\mathrm{P}}$, which for circular orbits is defined as the moment of maximum velocity of the primary star. We also initially set the eccentricity $e$ and argument of periastron $\omega$ as free parameters, but found that $e$ is indifferent from zero and kept it fixed in the final solution. The V2FIT also allows searching for the difference of systemic velocities of each component $\gamma_2-\gamma_1$, but we also found it to be insignificant and held it fixed to zero. Finally, we noted that there is a measurable difference in the zero points between HIDES and REOSC, and we initially set this as a free parameter as well. We assumed a common zero point for HIDES and HARPS, because with only one HARPS spectrum, the difference cannot be reliably measured, and the HARPS measurements did not deviate significantly from the HIDES measurements.

As a second approach, we used the PHysics Of Eclipsing BinariEs \citep[PHOEBE,][]{Prsa11} code (version 0.31a). The RV data were analysed with the program RADVEL \citep{Pribulla15}, which yielded the mass ratio and $\gamma$. These two parameters were fixed for the fitting procedure. Furthermore, the values of $a_{\mathrm{1,2}} \sin i$ gleaned from this program were taken as starting values. The eccentricity was fixed and considered equal to zero. The $T_{\mathrm{eff}}$ of the primary component was fixed to 12\,500\,K according to the results of profile fitting of the H$\alpha$ and H$\beta$ lines \citep{Niem17}. The $T_{\mathrm{eff}}$ of the secondary component was set to a starting value of 8\,000\,K. A logarithmic law for limb darkening was used, with coefficients interpolated from \citet{vanh93}. Because of the achieved precision of the photometric data, the selection of the limb-darkening law did not have any significant effect on the quality of the fit. Values of gravity brightening and albedos for both components were fixed to one because the effective temperatures indicate radiative envelopes for both components \citep{Claret99}. In order to test the validity of this approach, we have investigated the bearing of different values of gravity brightening on our calculations and find that they do not significantly affect our synthetic light curves. This is because even large changes in gravity brightening produce changes in the synthetic light curves that are two orders of magnitude below the precision of our photometric observations.

\begin{figure}
\begin{center}
\includegraphics[width=85mm, clip]{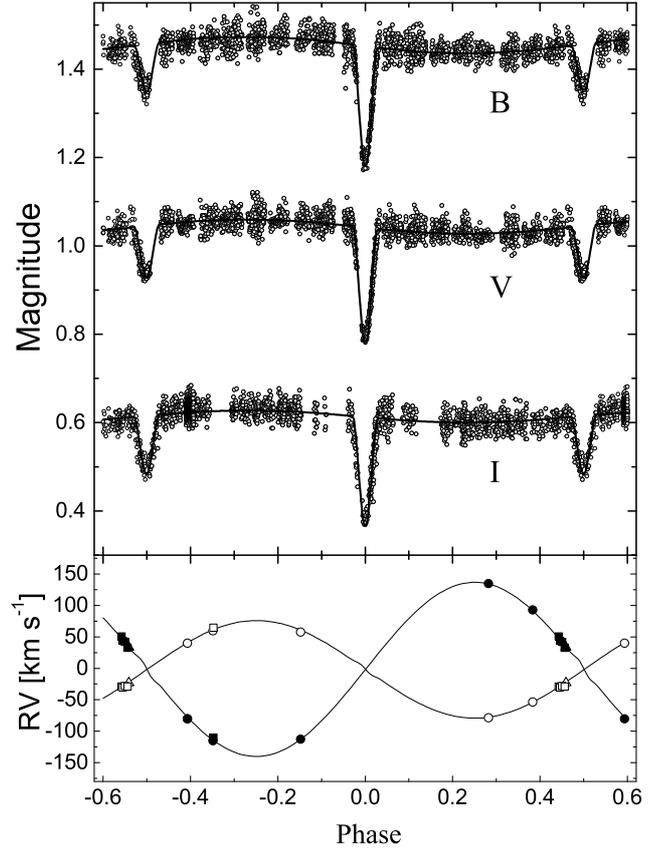}
\caption{Fit of the photometric (upper panel) and RV (lower panel) observations with the parameters listed in Table \ref{orbital_elements}.
The symbols correspond to the different instruments: circles (HIDES), triangles (HARPS), and squares (REOSC).}
\label{curves} 
\end{center} 
\end{figure}

\begin{figure}
\begin{center}
\includegraphics[width=85mm, clip]{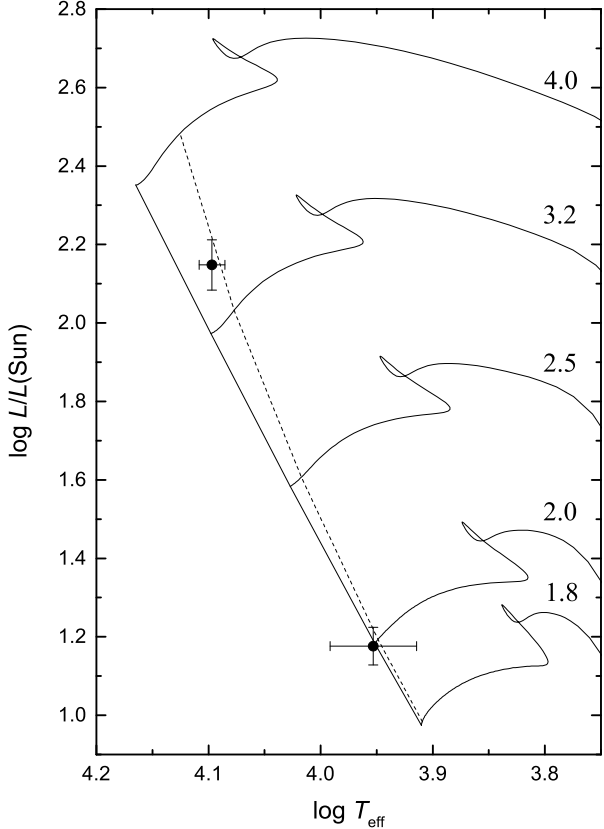}
\caption{Location of the two components in the \logTeff\ versus \logl\ diagram. The evolutionary tracks for different masses are taken from
\citet{Claret04}. The isochrone for 100\,Myr is indicated by the dotted line.}
\label{hrd} 
\end{center} 
\end{figure}

First, the values for the primary and secondary luminosity levels were set. After this, the surface potentials for both components along with the inclination ($i$), semi-major axis (SMA), and phase shift were fitted in order to match the eclipse widths of the synthetic and observed light curves. In an iterative process, the inclination and effective temperature of the secondary component were fitted to improve the quality of the fit around the secondary eclipse. The out-of-eclipse curvature of the phase curve with maximum amplitude near phase 0.25 indicates the presence of a large spot positioned perpendicular to the radius vector of the components.
The error estimation in PHOEBE is still somewhat problematic and tends to underestimate errors. The errors of the SMA, $i$, $T_{\mathrm{eff,2}}$ , and surface potentials were obtained directly from the differential correction method output. The errors of masses were calculated using $a_{\mathrm{1,2}} \sin i$ from RADVEL and the $i$ from PHOEBE.

We used the capability of PHOEBE to fit a spot to the light curve. Our results indicated two solutions with a cool spot on the primary or secondary component, and two 
solutions with a hot spot on the opposite sides of the components. After a first solution for each scenario, they were improved with the provided fitting methods. We were not able to improve the results by taking into account a spot on the secondary component. However, assuming a spot on the primary component, we were able to significantly improve both solutions, which converged to the same result, with the temperature differences ending up the same and the radii of the hot and cool spots being complementary to 180$\degr$. 
The derived temperature of the cool spot is 12\,235\,K assuming the 
effective temperature of the primary component as 12\,500\,K. We have to emphasize that the stellar spots of CP stars are due to
abundance inhomogeneities and not due to temperature differences as in solar-like stars. However, the effects for
the light curves are the same \citep{Koch17}. It is therefore not possible to directly correlate the temperature difference 
derived from PHOEBE to an abundance difference on the stellar surface.
The significant Si excess of the primary component implies an optically bright spot scenario as analogous to magnetic CP2 stars \citep{Oksala15}. This result is a further important proof that the primary CP3 star exhibits a spotty surface that is responsible for the detected photometric variability, as suggested by \citet{Niem17}.

\subsection{Evolutionary status of the system} \label{evol_stat}

As a final step, we determined the age of the HD\,66051 system using the evolutionary models by \citet{Claret04} for [Z]\,=\,0.02 and [X]\,=\,0.70. Within these models, the effective temperatures of the more massive stars are corrected for the effects of stellar winds. Convective core overshooting is assumed to be moderate and is modelled with $\alpha_{\mathrm ov}$\,=\,0.20. The models also take into account binary evolution and were successfully tested and applied to countless binary systems \citep{Eggleton17}.

We employed the spectroscopically derived astrophysical parameters of the two components and their errors (Table \ref{orbital_elements}). First, we used the \logTeff\ and $M_{\mathrm{Bol}}$ values to derive masses from the isochrones. For the calculation of the luminosities, the absolute bolometric magnitude of the Sun $M_{\rm Bol}(\odot)$\,=\,4.75\,mag \citep{Cayrel96} was used. Figure \ref{hrd} shows the location of the two components in the \logTeff\ versus \logl\ diagram, together with the evolutionary tracks for different masses and the isochrone for 100\,Myr. The derived masses for the components (3.2 and 2.0\,M$_{\odot}$ for the primary and secondary component, respectively) are in excellent agreement with the corresponding values derived from the orbital solution (cf. Table \ref{orbital_elements}). The system is very close to the zero-age main sequence and is not very much evolved. Within the error boxes, we find a maximum age of about 120\,Myr or about 15\% of the total lifetime on the main sequence.

\section{Discussion} \label{discussion}

When we compare the derived results with the star HD\,65949 \citep{Cowley10}, which is the only other star exhibiting similar chemical peculiarities as our target star, we find that HD\,66051 is significantly younger. However, as pointed out, the two stars share common elemental anomalies, which were interpreted in the light of the common diffusion theory \citep{Michaud76,Alecian11} by \citet{Niem17}. The basic idea is to test the diffusion timescale for certain elements with binary systems such as HD 66051, which allows placing tight constraints on age. To this end, we have to assume that the stars arrive on the main sequence with abundances that are well mixed. Chemical separation then takes place as a result of a time-dependent process. For instance, the deficiency of N and the overabundance of P should develop before the appearance of a significant Mn excess (which is not present 
in HD\,66051). Another point in case is the shared absence of the high Ga abundance typical of many CP3 stars. According to the speculations of \citet{Cowley10}, the Ga anomaly may be a pure diffusion anomaly. It therefore might not have had the time to develop in both stars.

On the other hand, \citet{Cowley10} have put forth the hypothesis that the composition of HD\,65949 might have been influenced by accretion of exotic r-processed material that was subsequently differentiated by atomic diffusion. The suggestion that mass transfer might (also) play a role in the development of chemical peculiarities (in particular in regard to CP3 star anomalies) has been recurring for a long time \citep{Wahl95}. Furthermore, binarity is thought to play a vital role in the development and understanding of CP3 star anomalies \citep{Schoeller10}. With its unique configuration, the system of HD\,66051 will allow an investigation of these theories and add constraints on what (if any) processes in addition to atomic diffusion are at work in the formation of CP3 star anomalies.

For further tests of the time-dependent diffusion models, we need more investigations of spectroscopic binary systems such as HD\,66051. This will allow us to add further constraints on the evolution of elemental abundances during the main-sequence lifetime of B-type stars in general. Hopefully, satellite-based missions such as Gaia \citep{Mowlavi17} and TESS \citep{Camp16} will provide accurate astrometric, kinematic and photometric data for these rare systems. Together with follow-up spectroscopic observations, a detailed analysis will shed more light on the ongoing processes in stellar atmospheres of stars of the upper main sequence.

\section*{Acknowledgments}
This project was supported by grant 7AMB14AT030 (M\v{S}MT) and also by a grant of the Slovak Research and Development Agency with the number APVV-15-0458PPP. KGH acknowledges support provided by the Polish National 
Science Center through grant no. 2016/21/B/ST9/01613. EN acknowledges the Polish National Science Centre grant no. 
2014/13/B/ST9/00902.

\label{lastpage}
\end{document}